\documentclass[linenumbers]{raa_rb}
\usepackage{graphicx,times}
\usepackage{natbib}
\usepackage{amssymb,amsmath}
\bibpunct{(}{)}{;}{a}{}{,}
\usepackage[pagebackref=true]{hyperref}
\usepackage{longtable}

\begin{document}

\title{Unveiling the initial conditions of open star cluster formation}

\volnopage{ {\bf 20XX} Vol.\ {\bf X} No. {\bf XX}, 000--000}
\setcounter{page}{1}

 \author{C. J. Hao\inst{1,2}, Y. Xu\inst{1,2}\footnote{Corresponding author}, L. G. Hou\inst{3}, 
 Z. H. Lin\inst{1,2}, Y. J. Li\inst{1}
   }

\institute{Purple Mountain Observatory, Chinese Academy of Sciences, 
Nanjing 210023, PR China; {\it xuye@pmo.ac.cn}\\
\and
School of Astronomy and Space Science, University of Science and Technology of China, 
Hefei 230026, PR China\\
\and
National Astronomical Observatories, Chinese Academy of Sciences, 20A Datun Road, 
Chaoyang District, Beijing 100101, PR China\\
\vs \no
{\small Received 20XX Month Day; accepted 20XX Month Day}
}

\abstract{Open clusters (OCs) are infrequent survivors of embedded clusters 
gestated in molecular clouds.
Up to now, little is known about the initial conditions for the formation of OCs.
Here, we studied this issue using high-precision astrometric parameters 
provided by {\it Gaia} data release 3.
The statistics show that the peculiar motion velocities of OCs vary little 
from infancy to old age, providing a remarkable opportunity to use OCs
to trace their progenitors. 
Adopting a dynamical method, we derived the masses of the progenitor clumps 
where OCs were born, which have statistical characteristics 
comparable to previously known results for clumps observed in the Galaxy.
Moreover, the masses of the progenitor clumps of OCs indicate they should be 
capable of gestating massive O-type stars.
In fact, after inspecting the observed OCs and O-type stars, we found that 
there are many O-type stars in OCs.
The destructive stellar feedback from O-type stars may disintegrate the vast 
majority of embedded clusters, and only those sufficiently dense ones can 
survive as OCs.
}

\authorrunning{C. J. Hao et al. }            
\titlerunning{Open cluster formation}  
\maketitle

%
\section{Introduction}           
\label{sec:introduction}

The vast majority of stars in the Milky Way is believed to form in clusters 
of dozens to thousands of members in molecular 
clouds~\citep[e.g.,][]{lada2003,bressert2010,megeath2016}.
The observations of young star-forming regions~\citep[e.g.,][]
{feigelson2013}, theory~\citep[e.g.,][]{mcKee2007,heyer2015}, and 
simulations~\citep[e.g.,][]{offner2009} all have pictured star formation 
as a turbulent, clumpy, and stochastic process.
To some extent, star formation in crowded environment can determine 
the properties of stars themselves, such as the initial mass function 
(IMF), and stellar multiplicity distributions~\citep{sills2018}.
However, understanding of the formation and evolution of stellar clusters
is still poor as these objects are deeply embedded in molecular clouds 
in their early evolutionary stages, and hence not optically observable, and 
new puzzling observations continuously challenge theoretical models, 
so they remain a fascinating topic today~\citep[e.g.,][]{krause2020}.
%

\begin{figure*}
\begin{center}
\includegraphics[width=0.85\textwidth]{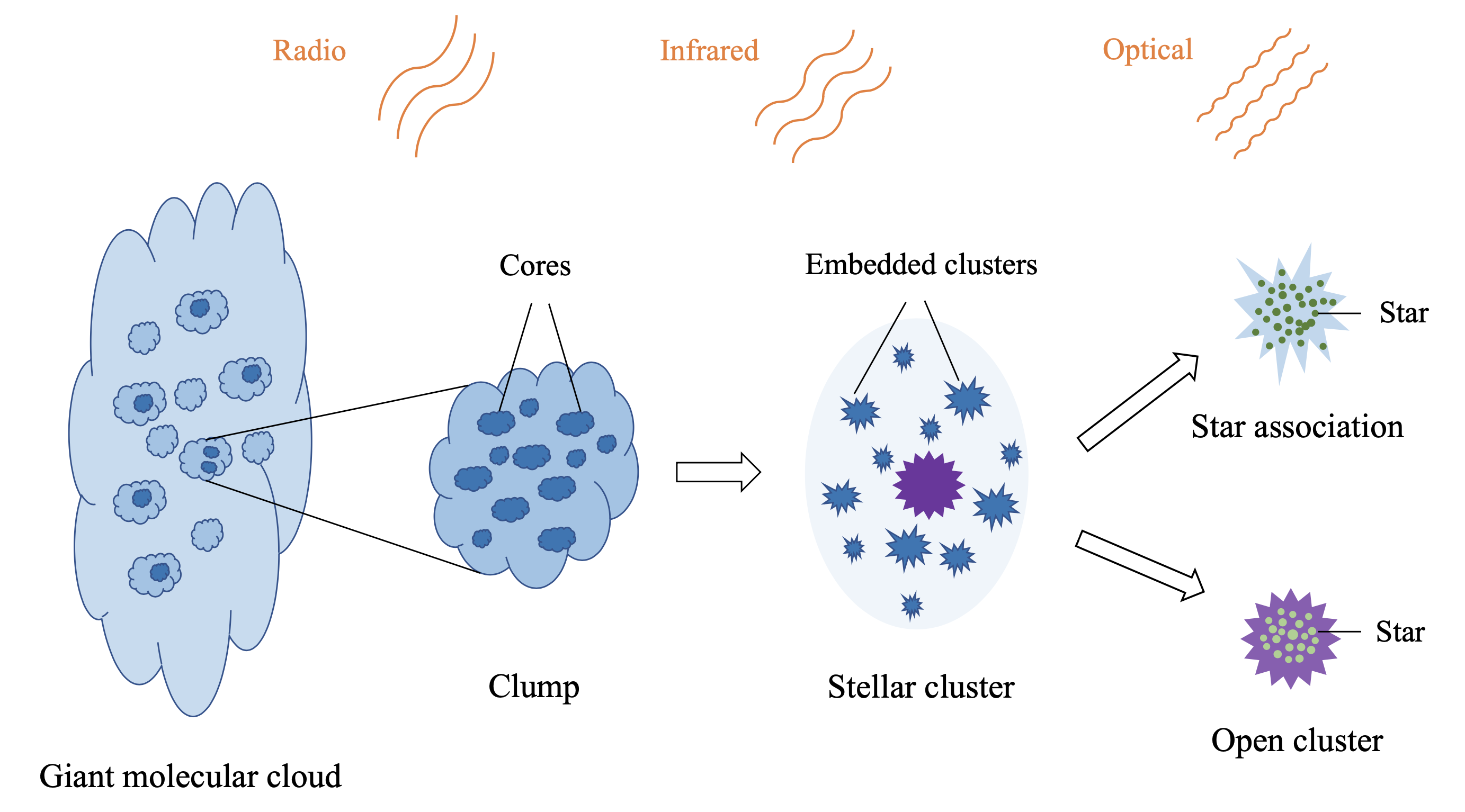}
\caption{Sketch map of the evolutionary pathway from clumps in a GMC 
to proto stellar clusters consisting of embedded clusters, and ultimately to 
bound open clusters and/or unbound star associations.}
\label{fig:fig1}
\end{center}
\end{figure*}

Figure~\ref{fig:fig1} presents the pathway from radio observed molecular clouds and/or
clumps, to proto stellar clusters consisting of embedded clusters that are often only 
visible at infrared wavelength, and ultimately to the optically identified star associations 
and/or OCs.
Giant molecular clouds (GMCs), as the vast assemblies of molecular gas, 
possess masses from $\sim$$10^{3}$ ${\rm M}_{\odot}$ to 
$\sim$$10^{7}$ ${\rm M}_{\odot}$~\citep[e.g.,][]{elmegreen1996,murray2011}. 
Galactic clumps, as the dense parts of GMCs, gestate many denser 
cores, which are the nurseries of embedded clusters~\citep[e.g.,][]
{lada2003,rathborne2006,mcMillan2007}.
It has become very clear that not all stars form in relaxed, centrally 
concentrated structure, and can often form in complex hierarchical or 
substructured distributions that follow the gas~\citep[e.g.,][]{whitmore1999,
schmeja2008,wright2014,krumholz2019}.
For example, the best studied embedded cluster, $Trapezium$, is within 
the more extended Orion Nebula Cluster~\citep{kuhn2019}.
However, it has been suggested that the vast majority of embedded
clusters will evolve into unbound star associations, and only a few 
percent (4--7\%) will survive as bound OCs~\citep[e.g.,]
[]{lada2003,bastian2006}, as illustrated in the sketch map shown in 
Figure~\ref{fig:fig1}.
On average, each GMC or GMC complex probably produce one bound 
open star cluster~\citep{elmegreen1985}, and stars in such systems 
account for about 10\% of all stars in our Galaxy~\citep{roberts1957,
adams2001}.
Although efforts in both observations~\citep[e.g.,][and references within]
{lada2003} and numerical simulations~\citep[e.g,][]{proszkow2009a,
proszkow2009b,girichidis2012,dale2015,farias2018} have been devoted 
to study the star formation and early evolution of embedded clusters, 
little is known about the initial conditions of OC formation.
The reason for the low survival rate of OCs arising from embedded 
clusters is still a mystery.

During the formation of stellar clusters, newborn stars could have 
profound effects on other stars and their natal molecular material, and 
many stellar feedback mechanisms would inject momentum into the 
star-forming environment~\citep{krumholz2014}, e.g., protostellar 
outflows~\citep{mckee1989,bally2016,li2020}, stellar radiation 
pressure~\citep{murray2010}, stellar winds from hot 
stars~\citep{vanKempen2010}, etc. Such stellar-feedback mechanisms 
are in principle enough to move all the surrounding 
material~\citep{krumholz2014}.
Indeed, the stellar system that forms in a clump may 
expand~\citep[e.g., Orion Nebula cluster,][]{kuhn2019} as it emerges 
from the molecular gas.
In this process, unlike other objects (e.g., binary or triple stellar systems 
and individual stars) whose kinetics can be changed easily, gravitationally 
bound OCs contain a large number of stars, making them potentially good 
kinematic fossils for investigating their progenitors.
The \textit{Gaia} mission has published its data release 3
\citep[\textit{Gaia} DR3,][]{gaia2016,gaia2022}, which includes astrometric 
and photometric measurements of about 1.8 billion stars of different types, 
ages and evolutionary stages, and the determinations of the radial velocities 
(RVs) of more than 33 million objects. Meanwhile, the data quality of 
\textit{Gaia} has been further improved.
On the other hand,
at present, thousands of OCs have been discovered in the Milky 
Way~\citep[e.g.,][]{hao2022,castro2022}, 
particularly with precise astrometric parameters~\citep[e.g.,][]{cantat2020a,
hao2021,tarricq2021}, which provide a good opportunity to investigate the 
characteristics of their progenitors.

The remaining paper is organized as follows. 
Section~\ref{sample} describes the sample of OCs used in this work.
The kinematic properties of OCs are studied in Sect.~\ref{pm},
which mainly concentrates on the peculiar motions of OCs in the Galaxy.
Then, adopting a dynamical method, we derived the masses 
of progenitor clumps where OCs were born in Sect.~\ref{mc},
and the statistical characteristics of derived clumps were also compared 
with the previously known results of Galactic clumps.
Next, we made an investigation in Sect.~\ref{ooc} to realize whether the 
present-day OCs house massive O-type stars, ultimately confirming 
the indication of derived progenitor clumps of OCs.
In Sect.~\ref{discussion}, we discussed the reason for the low survival rate 
of gravitationally bound OCs and explored which embedded clusters 
can evolve into long-lived OCs.
Finally, we summarized this work in Sect.~\ref{summary}.
%

\section{Sample}
\label{sample}


Up to now, thousands of OCs have been identified in \textit{Gaia} data,
and their ages cover a wide range, from a few million years (Myr) to 
billions of years.
Based on previous works~\citep[i.e.,][]{koposov2017,cantat2018,cantat2019,
castro2018,castro2019,castro2020,liu2019,ferreira2019,sim2019}, \cite{cantat2020a} 
determined the parameters of 2 017 OCs found in \textit{Gaia} data release 
2~\citep[\textit{Gaia} DR2,][]{gaia2018}.
Similarly, based on previous studies~\citep[i.e.,][]{dias2002,kharchenko2013,
dias2014,schmeja2014,scholz2015,castro2018,cantat2018,cantat2019,
castro2019,castro2020,liu2019,hao2020,ferreira2020,he2021},
\cite{hao2021} synthesized a sample of more than 3 700 OCs, whose parameters 
have been determined according to \textit{Gaia} 
early data release 3~\citep[\textit{Gaia} EDR3,][]{gaia2021}.

We compiled a large number of Galactic OCs with three-dimensional kinematic 
parameters through the following steps. For the OCs synthesized by \cite{hao2021}, 
after removing 134 potentially false positive or non-existing clusters reported in 
\cite{dias2002} and \cite{cantat2020b}, we cross-matched the 
remaining OCs with the 2 017 OCs listed in the work of \cite{cantat2020a}, 
where 1 821 non-repetitive OCs were found. Then, we cross-matched 
the members stars of 2 017 OCs compiled by \cite{cantat2020a} with the \textit{Gaia} 
DR3 data set and updated their astrometric parameters. Among these objects, 
there are 1 772 OCs that have member stars with RV measurements provided by \textit{Gaia} DR3. 
For the 1 821 OCs listed in \cite{hao2021}, we have also updated their astrometric 
parameters by using \textit{Gaia} DR3, and 1 456 OCs have member stars with RV 
measurements. Thus, 3 228 OCs with \textit{Gaia} RV measurements were obtained. 
For each of these OCs, we used a weighted procedure to determine its mean RV 
and RV uncertainty based on the errors of individual measurements, following 
\citet{soubiran2018}. 
In the end, after filtering 375 objects with RV uncertainties larger than 10 
km~s$^{-1}$, we gathered a sample of 2 853 OCs with 
reliable mean RV parameters.
Age parameters of the selected OCs come from \cite{cantat2020a} and \cite{hao2021}.
%
%

\begin{figure*}
\begin{center}
\includegraphics[width=0.96\textwidth]{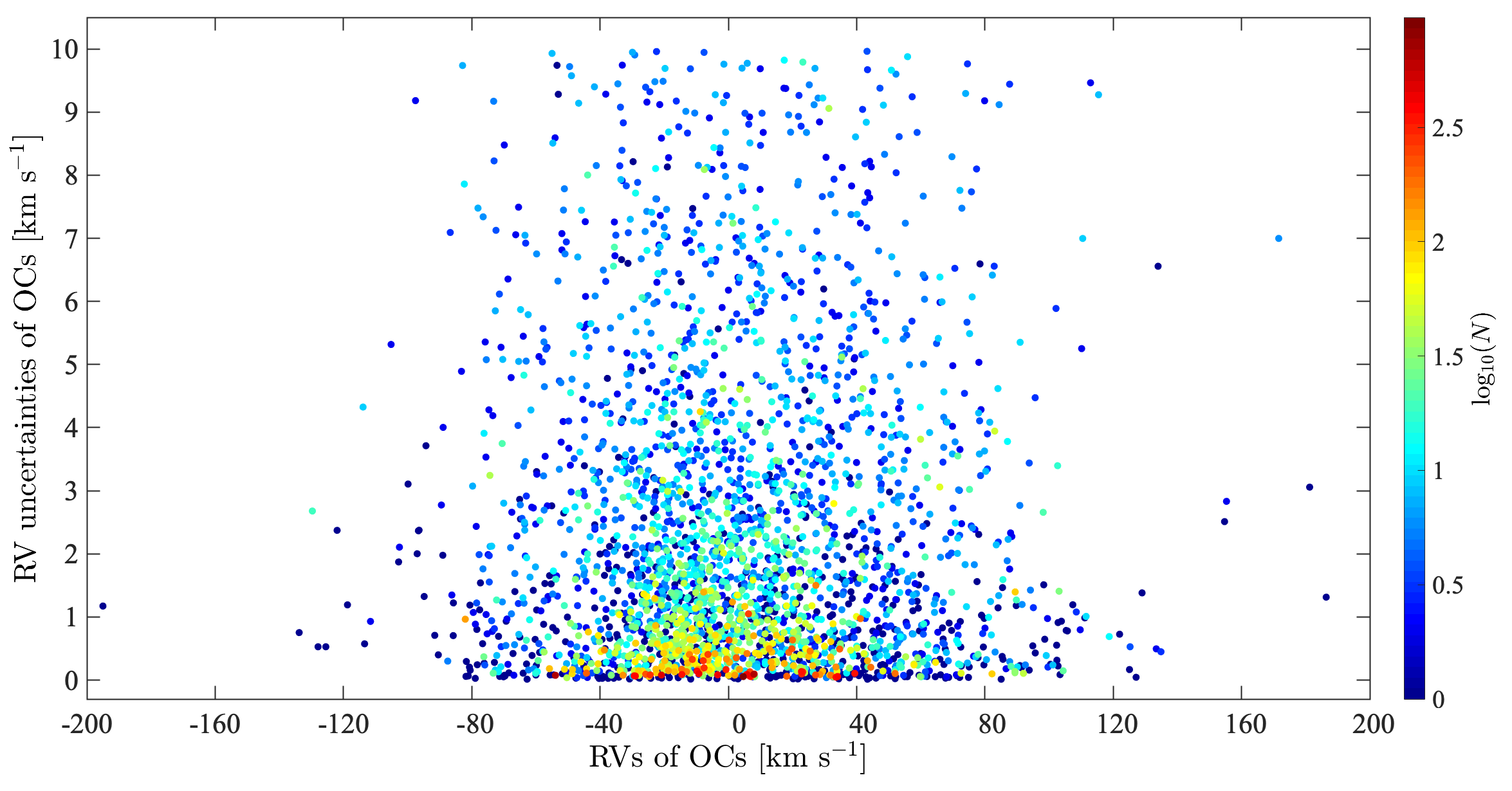}
\caption{RV uncertainties as a function of the RVs of OCs in the sample. 
The numbers ($N$) of OC member stars with RV measurements are colour coded.}
\label{fig:fig2}
\end{center}
\end{figure*}

The \textit{Gaia} DR3 data set is a large increase of the OC members that with 
RV measurements available.
Taking advantage of RV measurements from both \textit{Gaia} DR2 and ground-based 
spectroscopic surveys and catalogues, \cite{tarricq2021} computed the weighted RVs
and RV uncertainties of 1 382 OCs in \cite{cantat2020a}. 
Selecting the most reliable OCs that have an RV uncertainty lower than 
3~km~s$^{-1}$ based on at least 3 member stars, \cite{tarricq2021} obtained 513 
clusters in their sample.
Under this criterion, there are 1 317 OCs in our sample that can be considered to possess 
the most reliable mean RVs, 
which have a median RV uncertainty of 1.01~km~s$^{-1}$ 
and a median number of 14 member stars with RV measurements, benefiting from 
\textit{Gaia} DR3.
In Figure~\ref{fig:fig2}, we presented the RVs, RV uncertainties, and the numbers of 
member stars with RV measurements of 2 853 OCs in the sample.
For about 35\% (997 OCs) of the sample, the mean RV is based on more than 10 
member stars, and for about 71\% (2 015 OCs) it is based on at least 3 member stars. 
The RVs of 525 OCs are based on only one member star, which represent $\sim$18\% OCs 
of our sample.
2 415 OCs ($\sim$85\%) have RV uncertainties lower than 5~km~s$^{-1}$, and the 
RV uncertainties of 1 970 OCs ($\sim$70\%) are lower than 3~km~s$^{-1}$.
The median uncertainty of the weighted mean RV is 1.64~km~s$^{-1}$ when the 
full sample is considered.
The sample of 2 853 OCs was used for analysis in the next sections.
%

\section{Results}
\label{result}

\subsection{Peculiar motions of OCs}
\label{pm}

The large number of member stars of OCs makes them potentially 
good kinematic fossils for investigating their progenitors.
The peculiar motions (PMs) are non-circular motions with respect to 
the rotating Galactic disc and are significant kinematic 
attributes of OCs, the study of which enables to use OCs to trace their 
progenitors.
For the OCs obtained in Sect.~\ref{sample}, we have calculated 
their PM velocities ($v_{\rm pm}$), which were derived from 
their measured distances, proper motions, and radial velocities following
\cite{reid2009} and \cite{xu2013}.
In the Galactocentric reference frame, the three-dimensional motions
of OCs were straightforwardly calculated using the linear speeds
projected onto the celestial sphere.
Then, the PMs of OCs were estimated by subtracting Galactic rotation 
and the solar motions.
Here, a Galactic rotation speed near the solar circle of
236 $\pm$ 7~km~s$^{-1}$, a distance of the Sun to the Galactic centre
of 8.15 $\pm$ 0.15~kpc and solar motions of \textit{U}$_{\odot}$
= 10.6 $\pm$ 1.2 km~s$^{-1}$, \textit{V}$_{\odot}$ = 10.7 $\pm$
6.0~km~s$^{-1}$ and \textit{W}$_{\odot}$ = 7.6 $\pm$
0.7~km~s$^{-1}$ were adopted~\citep{reid2019}, where 
\textit{U}, \textit{V} and \textit{W} are the velocity components 
towards the Galactic centre, in the direction of Galactic rotation and 
towards the North Galactic Pole, respectively.
The PM velocities of OCs were defined as 
$v_{\rm pm}$ = $\sqrt{U^{2}+V^{2}+W^{2}}$.
%

\begin{figure*}
\begin{center}
\includegraphics[width=0.49\textwidth]{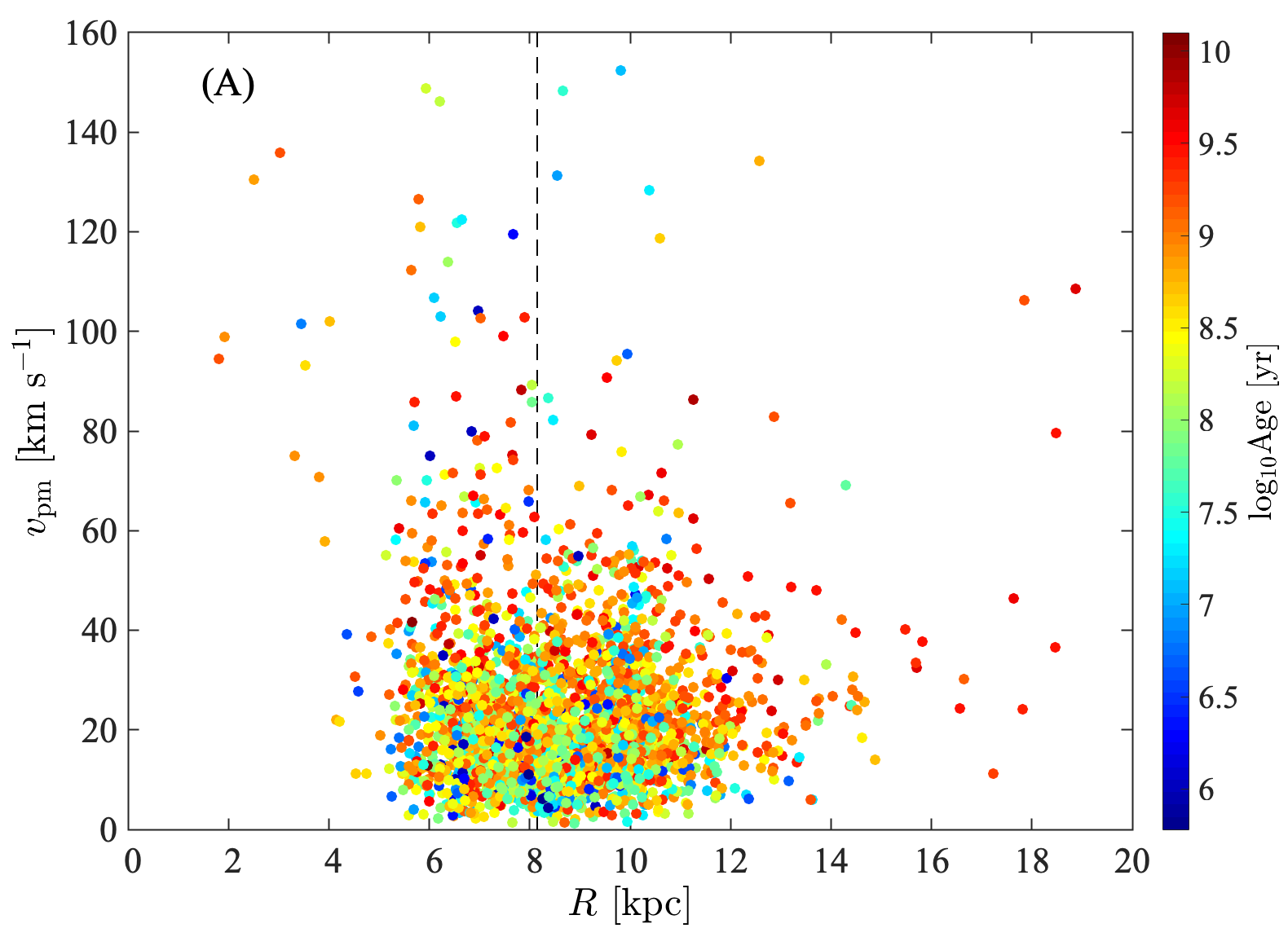}
\includegraphics[width=0.49\textwidth]{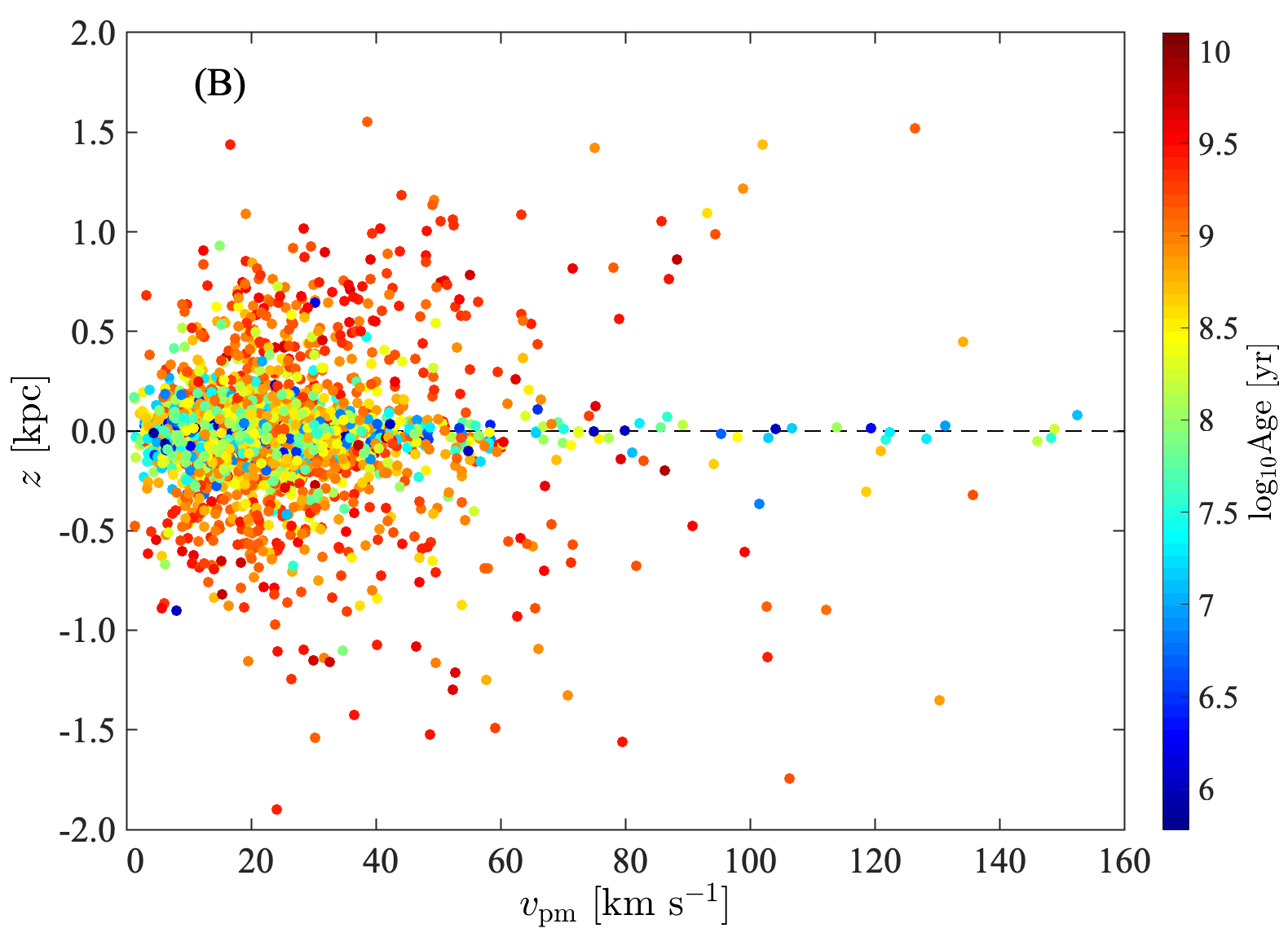}
\caption{Distributions of OCs with different PM velocities. {\it Panel} (A): OCs with different PM velocities as a function of Galactocentric distance. The Solar circle (black dashed line) is at 8.15 kpc~\citep{reid2019}. {\it Panel} (B): OCs with different PM velocities as a function of cluster $z$-height. The ages of the OCs are colour coded.}
\label{fig:fig3}
\end{center}
\end{figure*}

\begin{figure*}
\begin{center}
\includegraphics[width=0.49\textwidth]{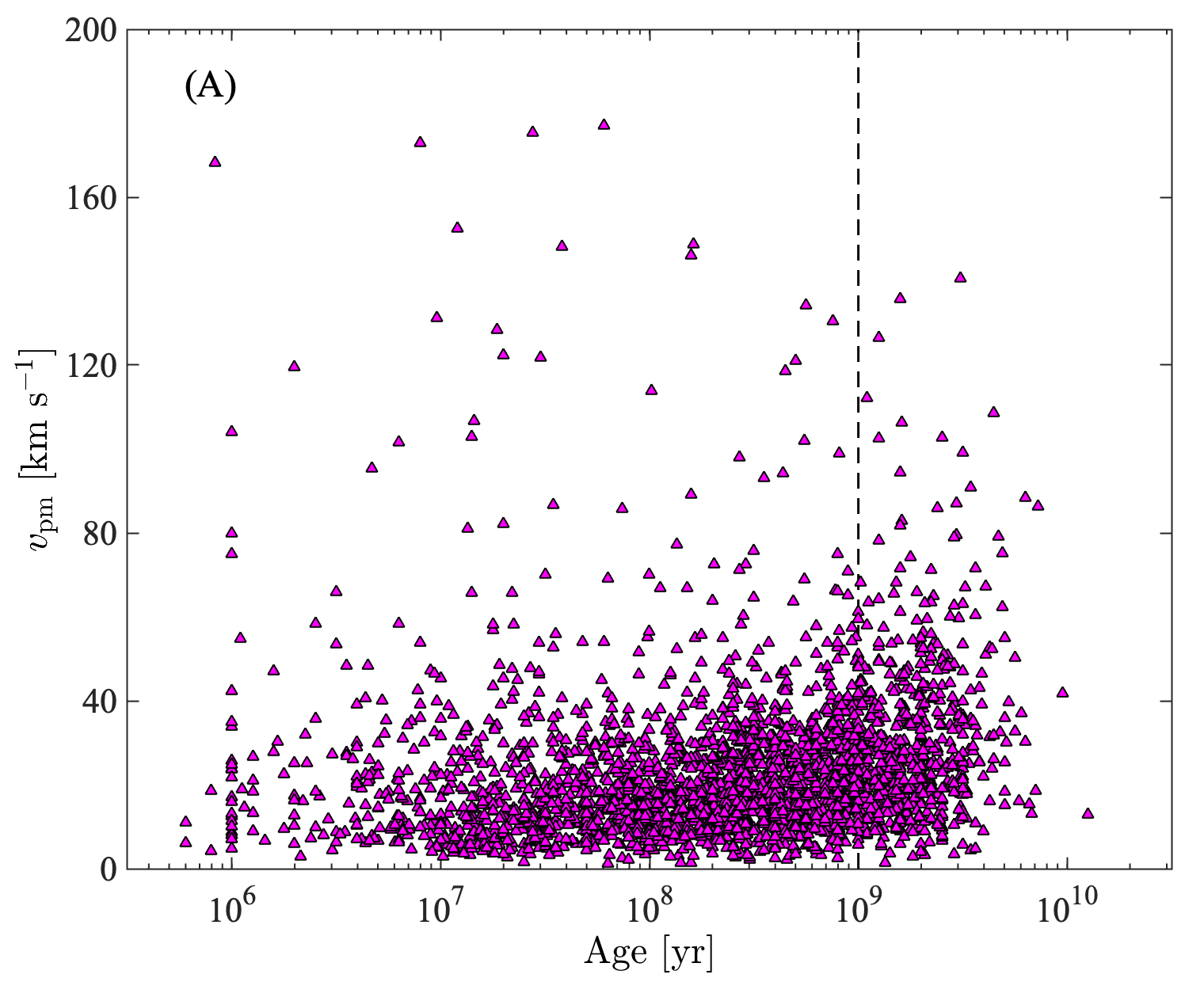}
\includegraphics[width=0.49\textwidth]{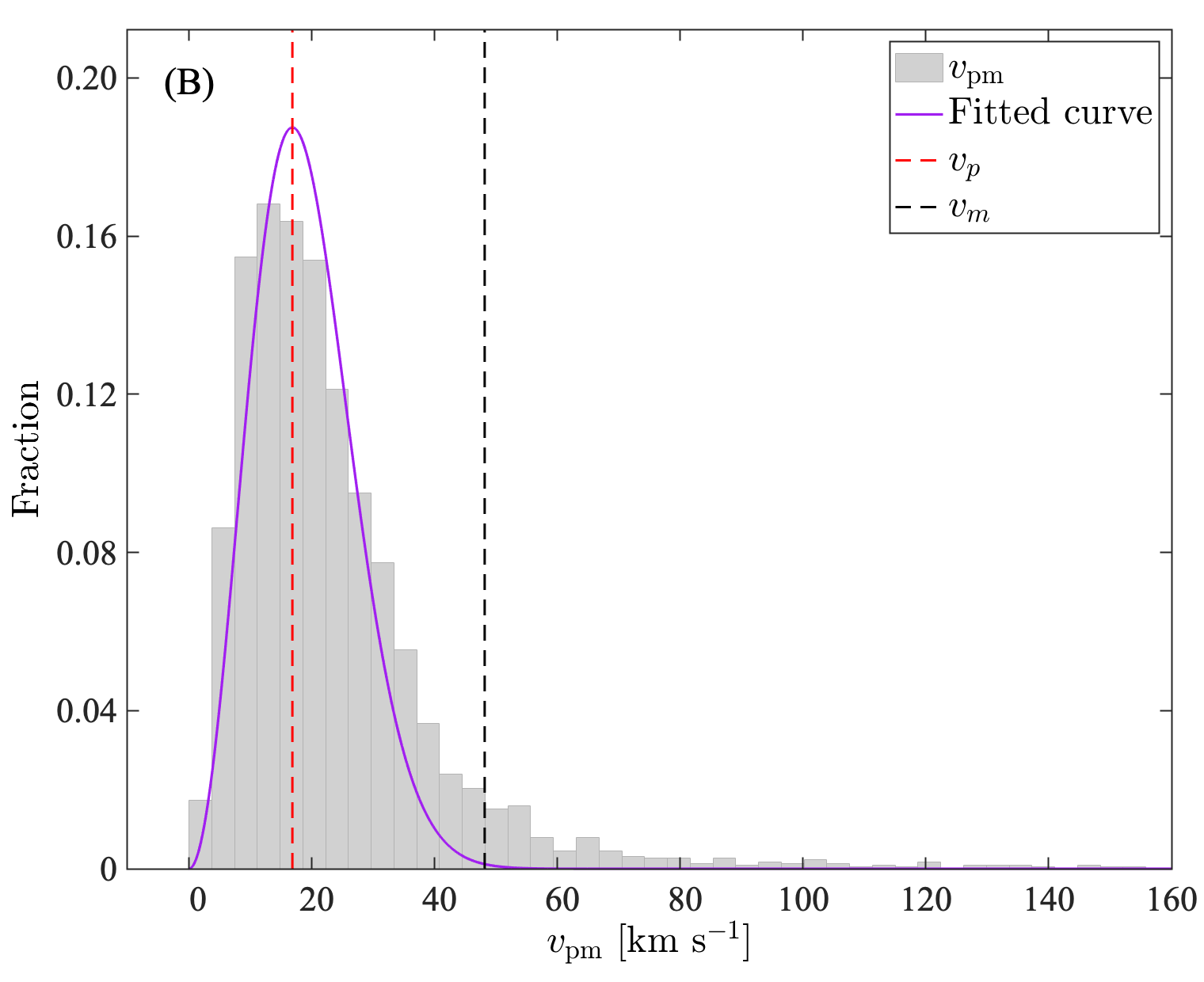}
\caption{Properties of the PM velocities of OCs. {\it Panel}
  (A): PM velocity as a function of cluster age. 
  The dashed black line is 10$^{9}$ years.
  {\it Panel} (B): distribution of the PM velocities of OCs. The solid purple line
  shows the best-fitting Maxwellian velocity distribution. 
  The dashed red line indicates the most probable velocity. 
  The dashed black line is the velocity of 48~km~s$^{-1}$.}
\label{fig:fig4}
\end{center}
\end{figure*}

Figure~\ref{fig:fig3}(A) displays the distribution of the 
PM velocities of OCs in different ages with respect to their 
Galactocentric distance, which demonstrates that the PM velocities 
of OCs at different distances from the Galactic centre are comparable.
Besides, almost all of OCs (99.2$\%$) are located between 
a Galactocentric distance range of [4, 16] kpc; hence, the influence of
the Galactic ``bar'' on the PM velocities of OCs in the sample should 
be negligible. 
Figure~\ref{fig:fig3}(B) shows the distribution of PM velocities of 
the OCs in different ages versus their $z$-heights from the Galactic 
middle plane,
which indicates that there is no significant distinction of the 
PM velocities of OCs with different $z$-heights.
Most OCs cross the Galactic plane several times in one orbital 
period and they gradually migrate from the Galactic
disk as they age \citep{wu2009,hao2021}.
The above results imply that there may be no difference in
the PM velocities of OCs when they travel in the Galaxy.

We made an investigation to address whether the PM velocities 
of OCs are variable as they age.
As shown in Figure~\ref{fig:fig4}(A), we present the PM velocities of 
OCs as a function of cluster age. 
For OCs younger than one thousand million years, there is no visible 
variation of the PM velocities with the increasing OC age.
The Pearson correlation coefficient (PCC), $\rho_{\rm X, Y}$, was 
used to evaluate the correlation between the PM velocities of the 
OCs and the cluster ages:
\begin{equation}
\begin{aligned}
\rho_{\rm X, Y} &= \frac{cov(X, Y)}{\sigma_{X}\sigma_{Y}} 
= \frac{E(X, Y) - E(X) E(Y)}{\sqrt{E(X^{2}) - E^{2}(X)}\sqrt{E(Y^{2}) - E^{2}(Y)}},
\end{aligned}
\end{equation}
where $cov$($X, Y$) denotes the covariance between the two 
variables, and $E$ is the mean of each variable.
The Pearson correlation coefficient (PCC) between the PM
velocities and ages is about 0.19 for all OCs in the sample, and only 
0.17 for OCs younger than 1 Gyr (Figure~\ref{fig:fig4}(A)).
Hence, the variation of PM velocities is small for OCs from
infancy to the old age of one thousand million years.

We also investigated the distribution characteristics of the PM 
velocities of OCs, as shown in Figure~\ref{fig:fig4}(B).
The distribution of the PM velocities of OCs can be fitted with a 
Maxwellian velocity distribution function:
\begin{equation}
f(v) = A \times e^{-B \cdot v_{\rm pm}^{2}} \cdot v_{\rm pm}^{2}.
\end{equation}
The best-fitting parameters of {\it A} and {\it B} are 4.06 and
0.0035, with 95$\%$ confidence intervals of [3.69, 4.44] 
and [0.0033, 0.0038], respectively.
According to the Maxwellian velocity distribution, the most probable
velocity is:
\begin{equation}
v_{p} = \sqrt{\frac{1}{B}}.
\end{equation}
Besides, the probability that the velocity is within the finite interval 
$[v_{\rm 1}, v_{\rm 2}]$ is:
\begin{equation}
P(v) = \int_{v_{1}}^{v_{2}} f(v)\, dv, \ \ \int_{0}^{\infty} f(v)\, dv = 1.
\end{equation}
Hence, it is shown that the probability of $v_{\rm pm}$ in the range
[0, 48]~km~s$^{-1}$~is 99.9\%~(see Figure~\ref{fig:fig4}(B)), and the 
most probable velocity of an OC, $v_{p}$, is 
$\sim$17~km~s$^{-1}$~(see Figure~\ref{fig:fig4}(B)).
Since the variation of the PM velocities 
is very small for OCs from infancy to the age of one thousand million 
years, it is likely to make a connection between the present-day 
OCs and their progenitors.
Then, the initial conditions for producing OCs can be revealed.
%


\subsection{The progenitor clumps of OCs}
\label{mc}

The present-day PM velocity of an OC has two possible origins: 
the separation velocity of the OC from the system where 
it born, or the inherited velocity from its natal system. 
The former scenario can be conceived that most embedded clusters 
evolve to be unbound star associations after separating from their natal 
systems, and ultimately become Galactic field stars, while only a few 
percent survive as bound OCs (see Figure~\ref{fig:fig1}), showing a 
very low fraction of bound OCs as announced in previous 
studies~\citep[e.g.,][]{lada2003,krumholz2019}. 
We found that there are several arguments can favor the former scenario, 
e.g., not all stars form in centrally concentrated but complex substructured 
distributions that follow the gas~\citep[e.g.,][]{wright2014};
stellar clusters formed in clumps are expanding when they 
emerging from the gas~\citep[e.g.,][]{kuhn2019};
the spatial distributions of gas and stars can determine whether the 
cluster remains bound or not~\citep[e.g.,][]{smith2011,smith2013}, 
and the substructured distribution of a stellar cluster can help it 
survive~\citep[e.g.,][]{allison2009}; etc.

Section~\ref{pm} shows that the PM velocities of OCs vary  
little from infancy to the old stage.
Thus, if we suppose that the PM velocities of OCs are nearly the
separation velocities from their natal systems, the masses of progenitors
that gave birth to OCs can be estimated,
which provides us a chance to study the initial properties of the OCs' progenitors.

Observations show that the dense clumps that are 
gravitationally bound have many denser cores~\citep[e.g.,][]{urquhart2014}. 
It has become clear that the stellar clusters formed in clumps are 
substructured, containing many embedded clusters born in denser cores, as 
illustrated in Figure~\ref{fig:fig1}. 
Here, similar to the definition given by \cite{kennicutt2012}, the scales 
(diameter) of clumps to which we refer are 1--10 pc and the 
denser cores are 0.1--1.0 pc.

The mass of a clump ($M_{\rm c}$) can be estimated as the total 
mass of stars ($M_{\star}$) and gas ($M_{\rm gas}$), where 
$M_{\rm c}$ = $M_{\star}$$/$SFE.
SFE, star formation efficiency, is a fundamental parameter of the 
star formation in a region, which is defined as:
\begin{equation}
{\rm SFE} = \frac{M_{\star}}{M_{\star} + M_{\rm gas}}.
\label{sfe}
\end{equation}
Here, $M_{\star}$ and $M_{\rm gas}$ are the total stellar and 
gaseous masses contained in the region, respectively.
The total potential energy of the stars before gas expulsion, 
$\Omega_{\rm 1}$, can be approximated by:
\begin{equation}
\Omega_{\rm 1} \sim - M_{\star} \cdot \frac{G \cdot M_{\rm c}}{r_{\rm h}},
\end{equation}
where $M_{\star}$ is the mass of stars, $r_{\rm h}$ is the
radius that contains half of the total mass in stars and $G$ 
is the gravitational constant. When gas is expelled, the 
potential energy of the system, $\Omega_{\rm 2}$, arises only from the 
stellar component, i.e.,
\begin{equation}
\Omega_{\rm 2} \sim - M_{\star} \cdot \frac{G \cdot M_{\star}}{r_{\rm h}}.
\end{equation}
The same as~\cite{farias2015,farias2018}, we assumed that the gas is expelled instantaneously.
Then, the stars have not had time to change their kinetic energy after gas 
expulsion, so we can assume $\Omega_{2} = {\rm SFE} 
\cdot \Omega_{1} $. Combining with Eq.~(\ref{sfe}), the separation 
velocity ($v$) of an OC from its natal proto-OC system can be approximated 
by the escape velocity as:
\begin{equation}
\begin{aligned}
v = \sqrt{ - \frac{2 \Omega_{2}}{M_{\star}}}
= \sqrt{ \frac{2GM_{\star}}{r_{\rm h}}} = \sqrt{ \frac{2GM_{\rm c} \cdot {\rm SFE}}{r_{\rm h}}}.
\label{vesc}
\end{aligned}
\end{equation}
In the following text, the term ``proto-OC system'' will refer to 
those proto stellar clusters formed in clumps, 
consisting of many substructures called embedded clusters. 
Then, the mass of a clump can be estimated as:
\begin{equation}
\begin{aligned}
M_{\rm c} = \frac{v^{2}r_{\rm h}}{2G \cdot {\rm SFE}}.
\label{mclump}
\end{aligned}
\end{equation}
If we want to derive the masses of progenitor clumps of  
OCs, there are three parameters need to be determined, i.e., the 
separation velocity $v$, the SFE, and the radius $r_{\rm h}$.

{\it Separation velocity $v$.}
The variation of the PM velocities of the OCs younger than one 
billion years is very small. 
For those OCs with ages of nearly one billion years, their  
PM velocity variations are estimated to be only about a few~km~s$^{-1}$.
In the following, we only selected the OCs younger than one billion 
years to do further statistical analyses and supposed their PM 
velocities are almost the separation velocities from their natal proto-OC 
systems, i.e., $v_{\rm pm}$ $\simeq$ $v$.
%

\begin{figure}
\begin{center}
\includegraphics[width=0.65\textwidth]{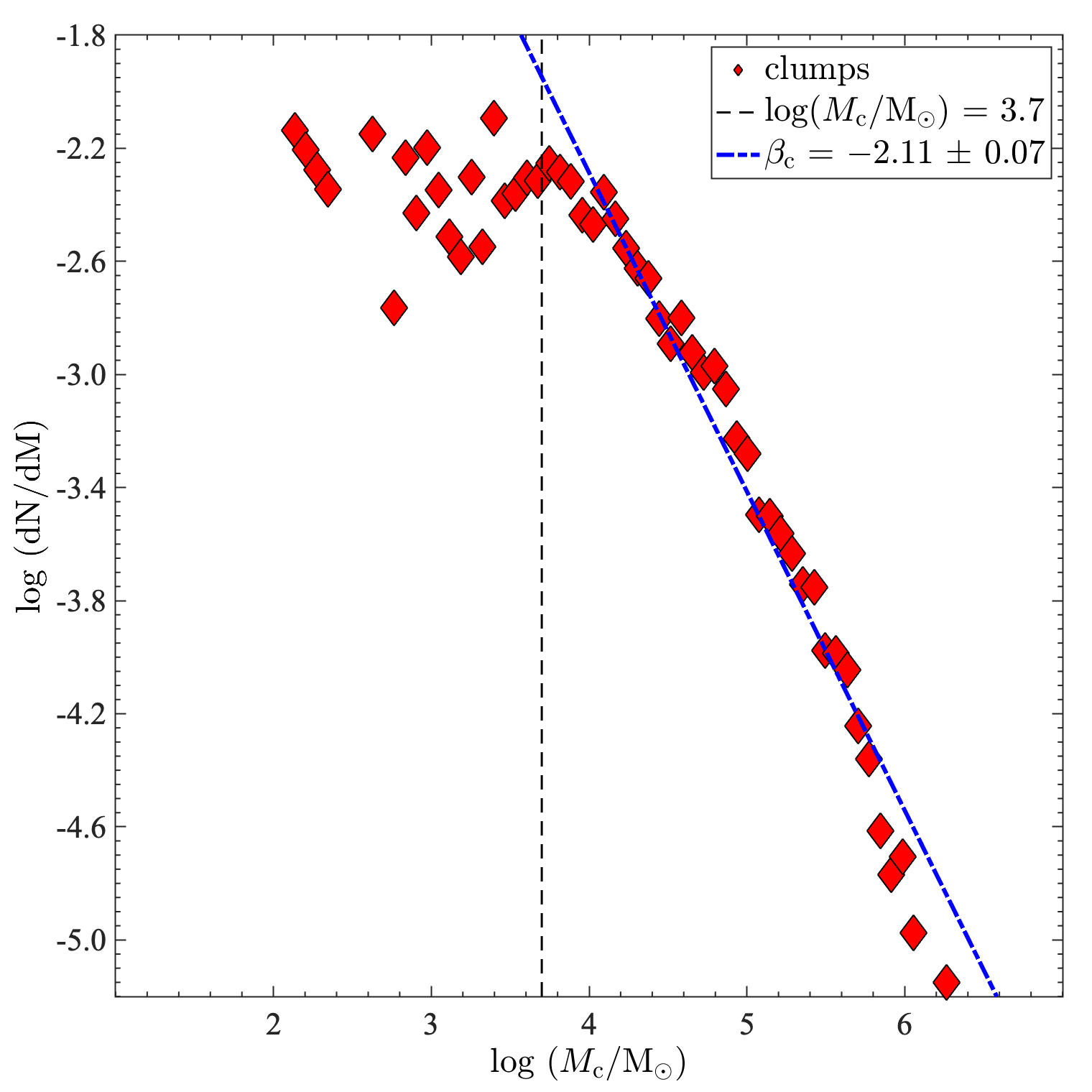}
\end{center}
\caption{Mass function of clumps that can produce
    OCs. The blue line indicates a maximum likelihood fit of a power
  law to the mass function, where the best-fitting index is $\beta_{\rm c}$
   = $-$2.11 $\pm$ 0.07. 
   The vertical dashed line shows the
  adopted lower-mass limit at log($M_{\rm c}$$/$${\rm M}_{\odot}$) =
  3.7.}
\label{fig:fig5}
\end{figure}

{\it SFE}.
Estimates of SFE are indirect and uncertain, e.g., the value of SFE 
globally observed for GMCs is 1--5$\%$~\citep{duerr1982,grudic2018}, 
and in star-forming regions of embedded clusters, SFEs range from 
approximately 10--30$\%$~\citep{lada2003}, while a bound OC would 
emerge only if SFE is greater than 50$\%$~\citep{wilking1983}. 
Considering not all embedded clusters in a proto-OC system 
can survive as bound OCs, we adopt SFE = 40$\%$ for 
the systems that can produce OCs.

{\it Radius $r_{\rm h}$}.
As mentioned above, stars form in the proto-OC system are 
substructured that follow the gas. While considering the stars 
are more concentrated than the gas~\citep[e.g.,][]{krumholz2019}, 
the $r_{\rm h}$ of proto-OC systems that we adopted are slightly 
smaller than the radii of clumps.
Referring to the clumps found in the submillimetre survey 
ATLASGAL~\citep[Atacama Pathfinder Experiment Telescope Large 
Area Survey of the Galaxy,][]{urquhart2014}, the radii, 
$r_{\rm h}$, were set to the range [0.3, 3.0] pc.
It can be expected that larger PM velocities of OCs implies 
richer and more massive progenitor clumps, because 
there are more significant momentum injection. There is a mass--radius 
relation for the Galactic clumps~\citep{krumholz2019}, i.e., 
$r$ $\propto$ $M^{\alpha}$. 
Combining this relation with Eq.(~\ref{vesc}), we can obtain $r$ 
$\propto$ $v^{2\alpha/(1-\alpha)}$. 
Index ($\alpha$) of the mass--radius relation for the 
clumps are in the range of 0.3--0.6~\citep{wong2008,romanduval2010,
urquhart2018}. 
Here, the adopted value of $\alpha$ is 0.5. 
Besides, we have chosen different values of $\alpha$ and the following 
results are not significantly different.

We first extracted OCs with ages younger than 1 Gyr. 
Then, since the fitted Maxwellian velocity distribution in Sect.~\ref{pm} 
shows that the probability of the PM velocities of OCs in the range 
[0, 48]~km~s$^{-1}$~is 99.9\%,
we rejected OCs with PM velocities larger than 48~km~s$^{-1}$, 
and the sources with PM velocity uncertainties larger than 10~km~s$^{-1}$ 
were also eliminated, eventually obtaining a subsample of 
1 571 OCs. 
For these OCs, the masses of their natal clumps ($M_{\rm c}$) are 
deduced with the above dynamical method, which
produces a range from $10^{2}$ ${\rm M}_{\odot}$ to $10^{6}$ 
${\rm M}_{\odot}$.
Actually, a vast majority of progenitor clumps have masses of $10^{3}$ to 
$10^{6}$ ${\rm M}_{\odot}$, and only $\sim$1\% of them are 
smaller than $10^{3}$~${\rm M}_{\odot}$.
The mass of the clump corresponding to the most probable velocity, $v_{p}$, 
is 2.6 $\times 10^{4}$ ${\rm M}_{\odot}$, consistent with that about 
48\% derived clumps have the order of mass magnitude of $10^{4}$ ${\rm M}_{\odot}$.
The derived masses of clumps are mainly ($\sim$82$\%$) 
in the range from $10^{4}$ ${\rm M}_{\odot}$ to $10^{6}$ ${\rm M}_{\odot}$,
which are comparable to the expectations of clump candidates 
where young massive stellar clusters are expected to be 
found~\citep[e.g.,][]{urquhart2018}, and actually, such systems are anticipated 
to yield OCs~\citep[e.g.,][]{lada2003}.

The mass function of the derived progenitor clumps (clump mass 
function, CMF) of OCs was also
determined and compared with those of previously reported results.
As shown in Figure~\ref{fig:fig5}, the mass function for these 
progenitor clumps, $\psi (M_{\rm c})$ $\equiv$ d$N$$/$d$M_{\rm c}$ 
$\propto$ $M_{\rm c}$$^{\beta_{\rm c}}$, was obtained, where the 
best-fitting power-law exponent was found to be $\beta_{\rm c}$ = 
$-$2.11 $\pm$ 0.07. 
Our adopted lower-mass limit is at log 
($M_{\rm c}$$/$${\rm M}_{\odot}$) = 3.7, because below this limit, the 
mass functions begin to fall significantly deviate from the extrapolated 
power law. Next, we derived the best-fitting value of $\beta_{\rm c}$ and 
its error from a least-squares estimation of clump mass above the 
mass-fitting limit. In order to obtain the parameter index of the clump mass 
function, we fixed the bin widths (mass) and counted the number of clumps 
per bin. Besides, we adopted different values of $\alpha$ in the range 
[0.3, 0.6], and the resulting indices of $\beta_{\rm c}$ were within the 
uncertainty of above result.
The derived $\beta_{\rm c}$ is in good agreement with the value of
$-$2.12 $\pm$ 0.15 reported in the $Herschel$ InfraRed Galactic Plane
Survey~\citep{olmi2018}, commensurate with the result of $-$2.10 deduced 
from numerical simulations~\citep{guszejnov2015}, and slightly flatter than
$Salpeter$'s value \citep[$-$2.35,][]{salpeter1955}.
This value also indicates that the power-law exponent of clumps
harbouring predecessor OCs does not present a significant difference from
the overall sample of Galactic clumps.
Both the masses and mass function of the derived progenitor  
clumps of OCs are almost concordant with the previously reported results
of Galactic clumps, 
suggesting that the dynamic method adopted here should be reasonable, 
which also indicates a potential connection between the PM velocities of 
OCs and their natal clumps.
%


\subsection{OCs and O-type stars}
\label{ooc}

Massive O-type stars are believed to play a significant role in the 
formation and evolution of stellar clusters, and also have profound 
effects on open star cluster formation~\citep[e.g.,][]{lada2003}.
It is therefore of great interest to investigate whether there are
O-type stars presenting in OCs.

Stellar clusters with masses of a few hundreds
${\rm M}_{\odot}$ to $10^{5}$ ${\rm M}_{\odot}$ are expected to
contain more than one O-type star~\citep{weidner2013}.
Consequently, as described in Sect.~\ref{mc},
since the masses of progenitor clumps 
that can produce OCs are in the range of $10^{3}$--$10^{6}$ ${\rm M}_{\odot}$, 
they are massive enough to give birth to O-type stars.
The low-mass embedded clusters in the progenitor 
clumps generally are difficult to evolve into OCs~\citep[e.g.,][]{lada2003}, while 
for the high-mass embedded clusters, some of them probably have survived 
as OCs and still harbor O-type stars.
Therefore, we investigated the OCs in the sample to address whether they 
contain O-type stars or not.

\begin{figure*}
\begin{center}
\includegraphics[width=0.49\textwidth]{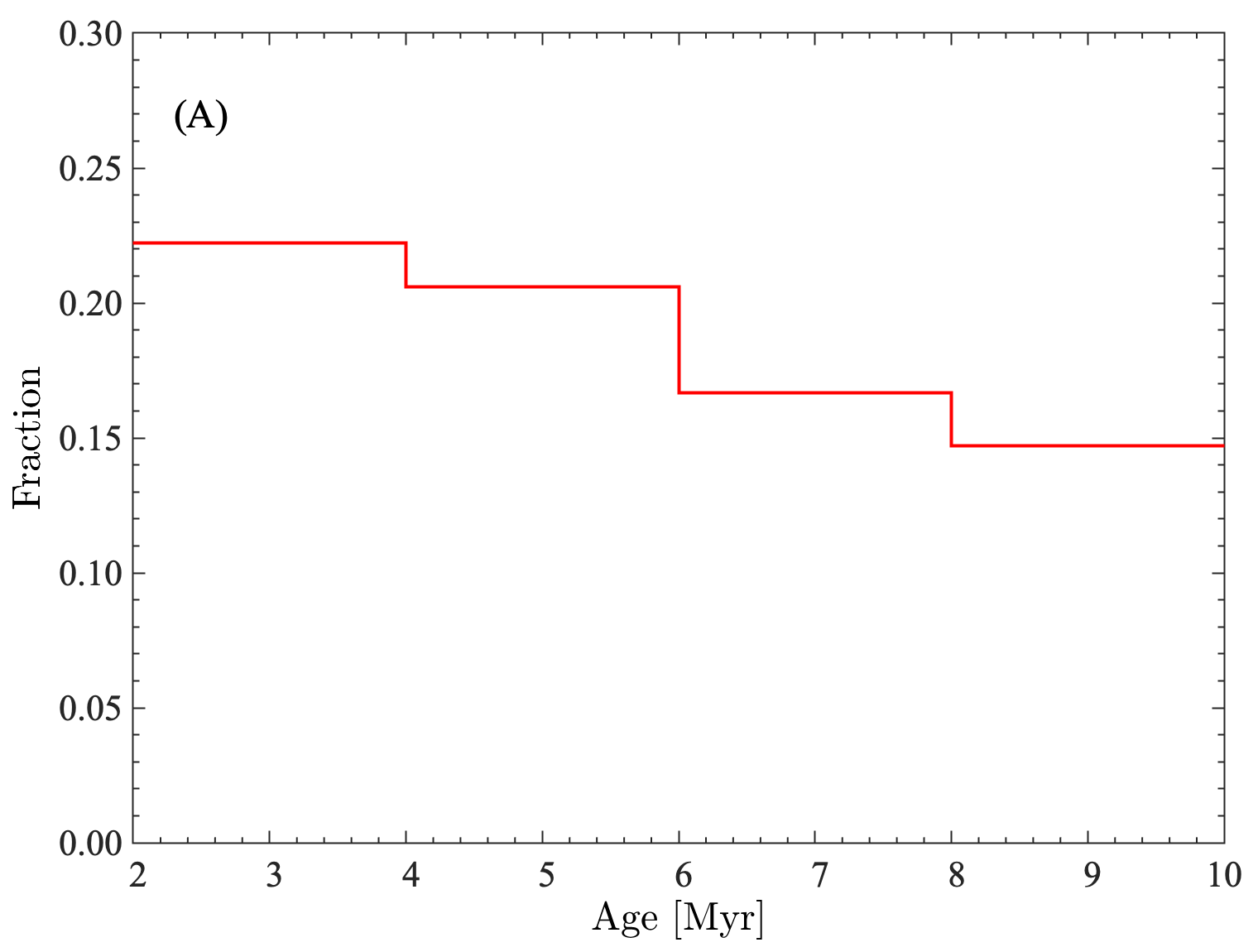}
\includegraphics[width=0.49\textwidth]{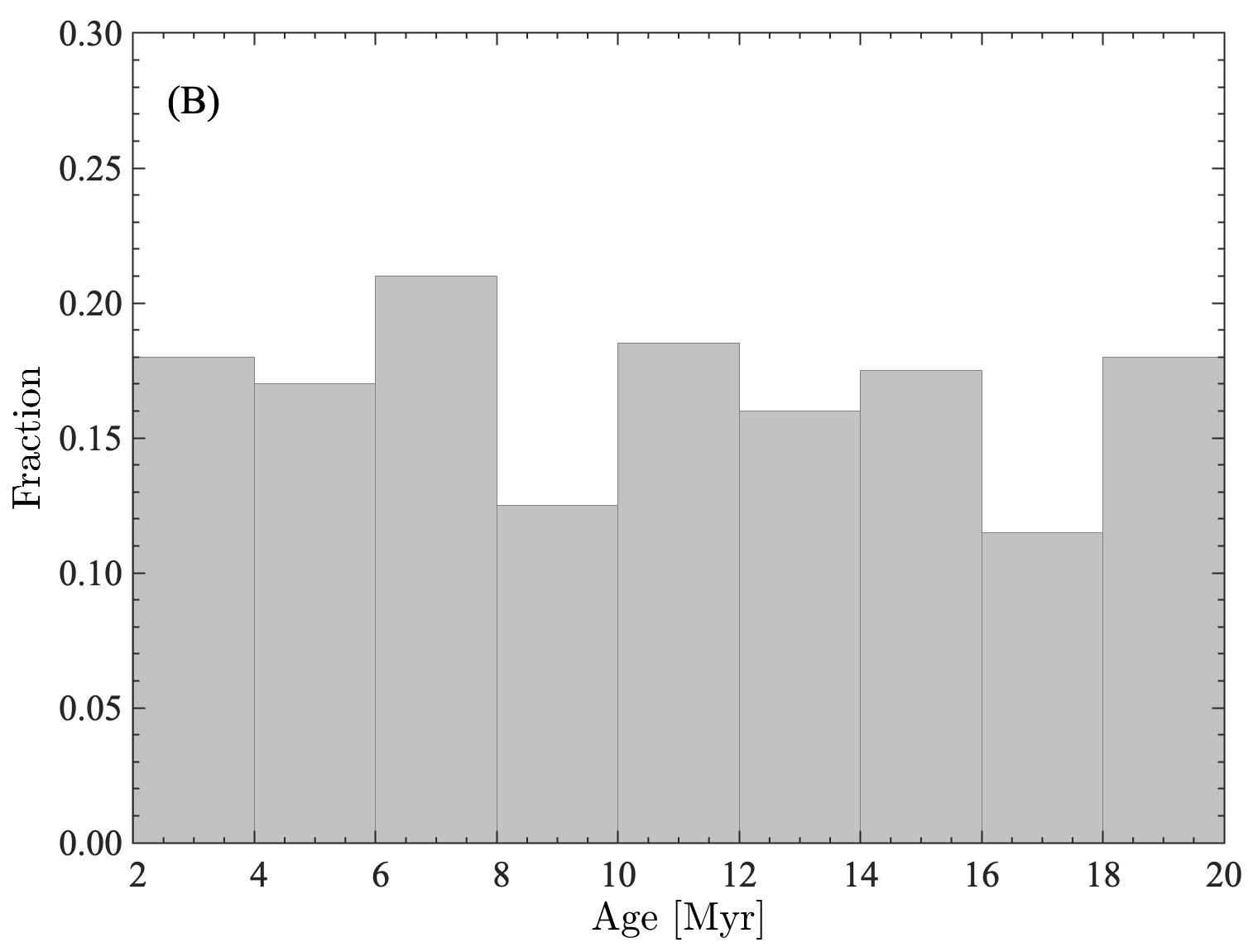}
\end{center}
\caption{OCs and O-type stars. {\it Panel} (A): fraction of
  OCs harbouring O-type stars for different age groups of OCs. {\it
    Panel} (B): fraction of OCs (2--20 Myr) in different age groups.}
\label{fig:fig6}
\end{figure*}

After inspecting OCs and the observed O-type stars, 
we found that there are many O-type stars present in present-day OCs.
The O-type star catalogue used in this work, containing 1 089 
O-type stars, was taken from \cite{xu2021}, who cross-matched 
the spectroscopically confirmed O-type stars collected by 
\cite{skiff2014} in \textit{Gaia} EDR3. After cross-matching the \emph{Gaia} 
\texttt{source\_id} of 
1 089 O-type stars with the 284 889 OC members in our sample, a 
total of 112 O-type stars were found in 56 OCs. 
Table~\ref{table:table_s1} in the Appendix~\ref{app1} 
presents these OCs, 
including their name, number and spectral types of O-type stars.
The fraction of young OCs ($\textless$ 10~Myr) harbouring massive
O-type stars is $\sim$18$\%$.
Especially, as shown in Figure~\ref{fig:fig6}(A), for OCs with ages of 
2 to 4 Myr, the fraction of OCs harbouring O-type stars is as high as   
22$\%$, which decreases to about 15$\%$ for OCs of 8 to 10 Myr.
%

\begin{figure}
\begin{center}
\includegraphics[width=0.72\textwidth]{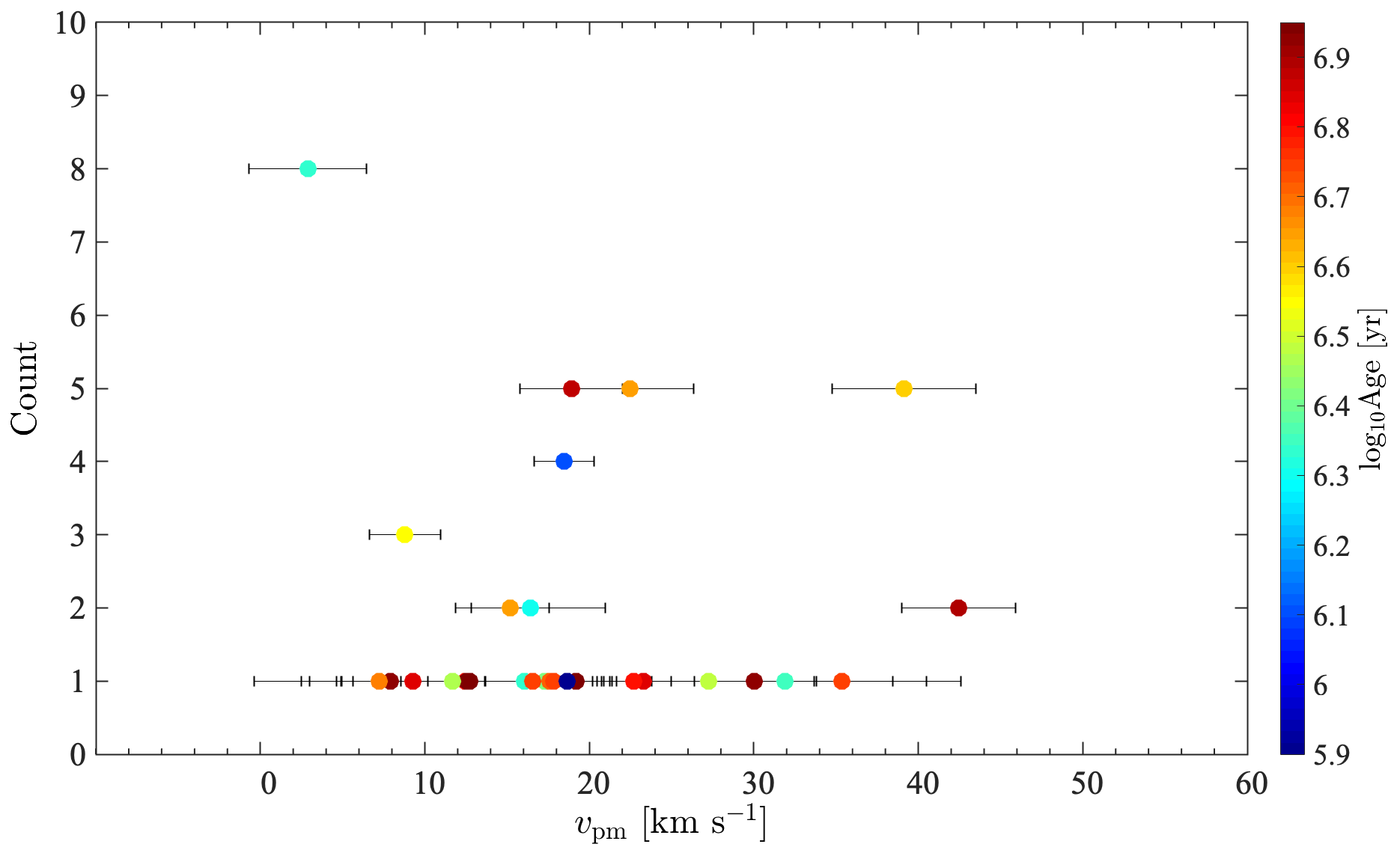}
\caption{PM velocities and errors of OCs ($\textless$ 10~Myr) containing O-type stars as a function of the number of O-type stars in the cluster. The error bars indicate the PM velocity uncertainties of the OCs. The ages of the OCs are colour coded.}
\label{fig:fig7}
\end{center}
\end{figure}

O-type stars are the most massive stars on the main sequence, 
and even the least massive O-type star has an initial mass of 16 
${\rm M}_{\odot}$~\citep{meynet2003}.
The most massive O-type stars spend less than one Myr on the 
main sequence and explode as a supernova after 3 or 4 Myr, while
the least massive ones can remain on the main sequence for about 
10 Myr, but cool slowly during this time and become early B-type 
stars~\citep{weidner2010}.
Thus, if an OC contained any O-type star, there is an upper limit 
on the age of the cluster, i.e., at least younger than 10 Myr.
As shown in Figure~\ref{fig:fig6}(B), the number of observed OCs 
does not obviously decrease at 3--4 Myr (i.e., the supernovae 
explosion timescale) and 10 Myr (i.e., the maximum lifetime of O-type 
stars), which suggests that the evolution of O-type stars probably 
not destroy their resident OCs.

We also studied the characteristics of the PM velocities of the OCs
harbouring O-type stars.
The median value of $v_{\rm pm}$ for young OCs (ages
$<$ 10~Myr) harbouring O-type stars is 
18 $\pm$ 3~km~s$^{-1}$,
which is similar to 
that of young OCs without O-type stars, i.e., 
17 $\pm$ 5~km~s$^{-1}$.
The mean $v_{\rm pm}$ of the young OCs containing O-type 
stars is 19~km~s$^{-1}$, while the corresponding values of OCs without 
O-type stars is 23~km~s$^{-1}$.
Figure~\ref{fig:fig7} shows the PM velocities of OCs containing O-type 
stars as a function of the number of O-type stars in the cluster.
For some (e.g., distant) OCs, the astrometric uncertainties translate into large 
PM velocity uncertainties.
The NGC 3603 and FSR 0696 OCs have significantly large PM velocity
uncertainties of 183~km~s$^{-1}$ and 78~km~s$^{-1}$, respectively, as their
relatively parallax errors are as high as 30\%.
Hence, the two clusters are not presented in Figure~\ref{fig:fig7}.
We found that the $v_{\rm pm}$ of OCs containing 1--2 O-type
stars are comparable to those with 5--8 O-type stars, indicating 
that there may be no relationship between the number of the harboured 
O-type stars and the PM velocities of OCs.
Besides, about 61$\%$ of young OCs harbouring O-type stars are 
located in the inner Galaxy, probably due to the presence of more 
numerous massive GMCs~\citep{heyer2015}.


\begin{figure*}
\begin{center}
\includegraphics[width=0.49\textwidth]{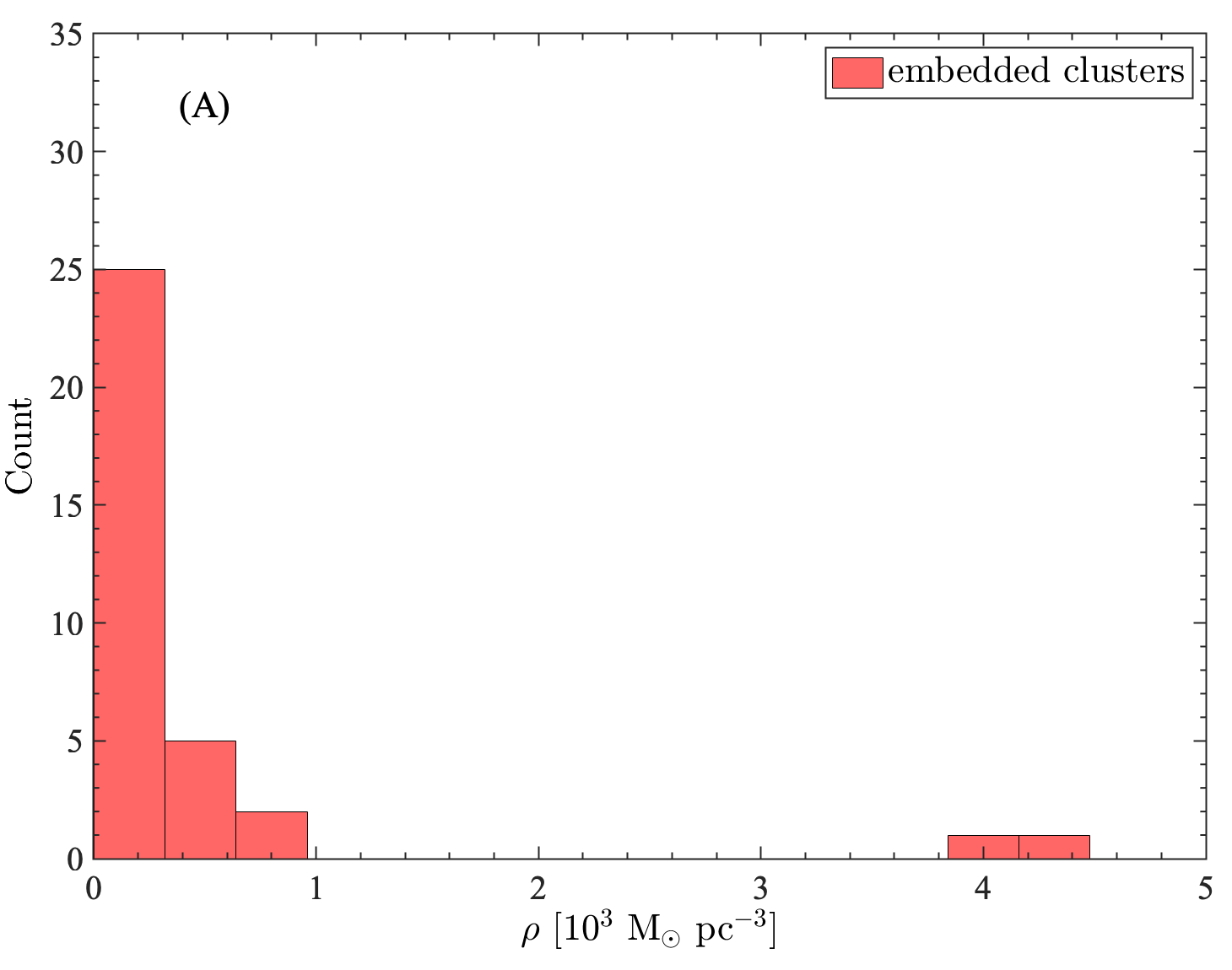}
\includegraphics[width=0.49\textwidth]{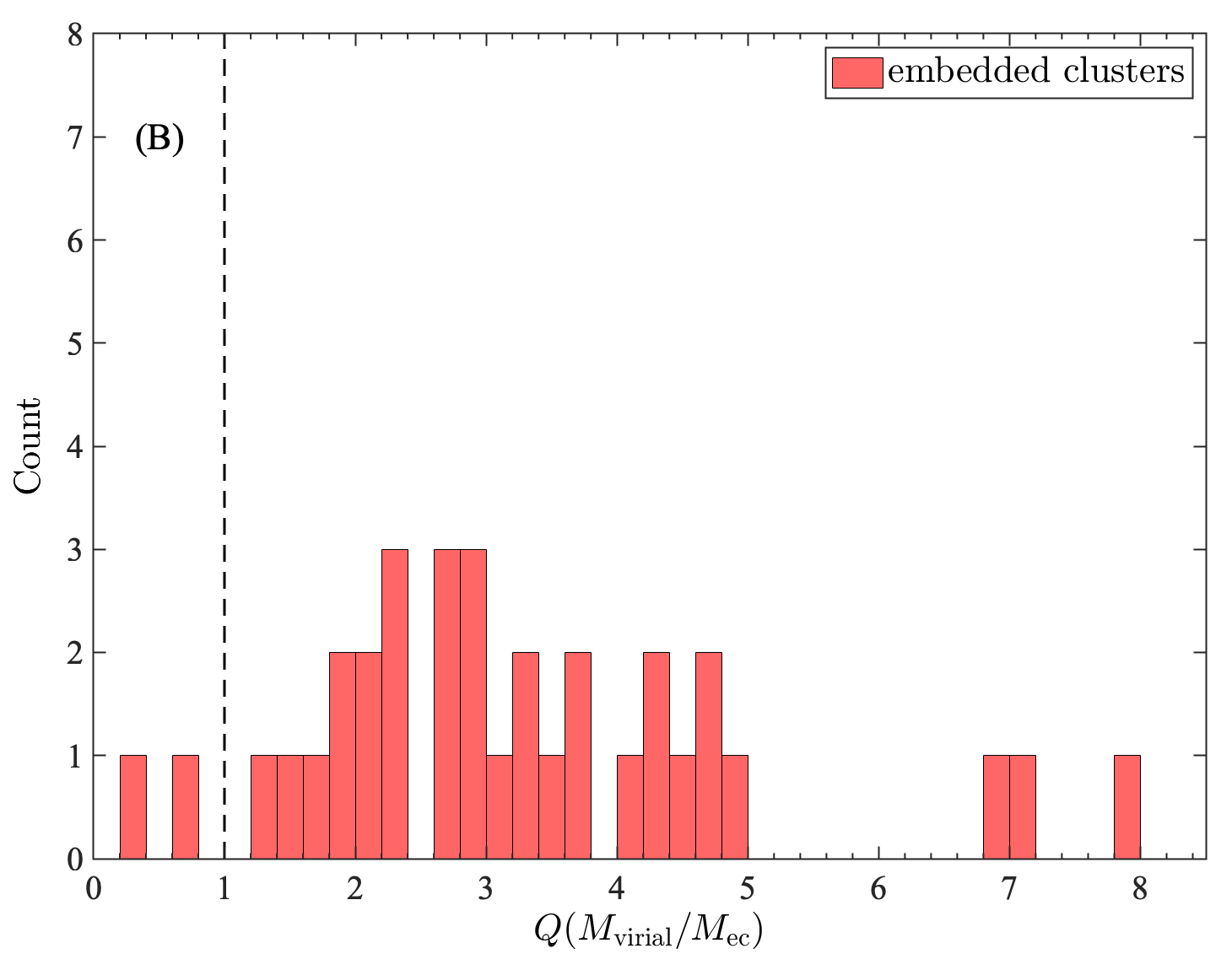}
\caption{Stellar mass density ({\it Panel} A) and virial parameter ({\it Panel} B) of known embedded clusters.
The dashed black line in {\it Panel} B is the virial parameter $Q$ = 1.}
\label{fig:fig8}
\end{center}
\end{figure*}

\section{Discussion}
\label{discussion}

The low fraction of gravitationally bound open star clusters
is still a mystery.
Previous studies have shown that various stellar feedback 
mechanisms play important roles in the formation and evolution of stellar 
clusters~\citep[e.g.,][]{mckee1989,kroupa2002,murray2010,
vanKempen2010,krumholz2014,bally2016,li2020}. 
As the most massive stars, the formation and evolution of O-type stars are 
accompanied by violent feedback to their surroundings in the form of 
copious amounts of ultraviolet radiation, powerful stellar winds and 
supernova explosions~\citep[e.g.,][]{dale2013,dale2008,dekel2013}.
Such destructive mechanisms can disperse the dense molecular material, 
and impede the birth of new stars, making it very difficult to satisfy the 
star formation efficiency~\citep[i.e., SFE $>$ 50$\%$,][]{wilking1983} 
needed for the formation of bound OCs from embedded 
clusters~\citep{lada2003}.
The results in Sect.~\ref{mc} and Sect.~\ref{ooc} have indicated that 
the progenitor clumps of OCs are capable of 
gestating O-type stars,
and particularly, many O-type stars are even present in present-day OCs.
The considerable influence caused by O-type stars in the
progenitor clumps of OCs probably results in a vast majority 
of embedded clusters can not survive and evolve into OCs.
Besides, the observed low SFEs ($<$ 50$\%$) for most embedded
clusters~\citep{lada2003} are probably also partially attributed to the 
existence of O-type stars.
However, this then raises a new specific question on which embedded 
clusters can survive in the conditions caused by violent feedback from 
O-type stars.

To study which embedded clusters can survive as bound OCs, we 
have conducted an investigation on the 
density properties of observed embedded clusters.
The stellar mass density, $\rho$, of 34 known embedded clusters was 
calculated based on their sizes and mass provided by \cite{lada2003}. 
The result was presented in Figure~\ref{fig:fig8}(A).
What is striking is that $\sim$6$\%$ embedded clusters
have a high stellar mass density of $\sim$4.0 $\times$ $10^{3}$ ${\rm
M}_{\odot}$ ${\rm pc}^{-3}$; in contrast, the others are all below
1.0 $\times$ $10^{3}$ ${\rm M}_{\odot}$ ${\rm pc}^{-3}$.
$Trapezium$, as one of the 6$\%$ of known embedded clusters with 
a sufficient stellar mass density, contains O-type stars, and has been 
identified as a possible predecessor of an OC~\citep{kroupa2001}.
Besides, it is interesting that the percentage of $\sim$6$\%$ is consistent 
with that of only 4--7\% of embedded clusters surviving as a bound
OC~\citep{lada2003}.

A further step has also been made to judge wether the observed embedded 
clusters would survive as OCs by estimating their virial parameter, 
$Q$ ($M_{\rm virial}/M_{\rm ec}$), as shown in Figure~\ref{fig:fig8}(B).
The same as \cite{krumholz2019}, $Q$ was determined as $Q$ $\equiv$ 
$5\sigma^{2}R/GM$, where $R$ is the radius and 
$M$ is the mass of the cluster. Here, we adopted a one-dimensional 
velocity dispersion, $\sigma$, of 0.7 km~s$^{-1}$, which is the typical 
limit value of observed young bound OCs~\citep{cantat2020b}.
Statistically, we found that only those $\sim$6 $\%$ of dense
embedded clusters have virial parameters of $Q$ $\textless$ 1,
which supports that they will likely evolve into bound stellar systems.

Conservatively, we speculated that the embedded 
clusters can survive as bound OCs as long as their stellar mass densities 
are sufficiently high.
However, it should be noted that the mass density here is 
a roughly mass density threshold for embedded clusters that can evolve 
to the phase of bound OCs.
The precise threshold of the mass density is expected to be determined 
using a further larger sample of embedded clusters.
%


\section{Summary}
\label{summary}
We conducted a pilot study on the formation of OCs in the Milky Way.
From infancy to the old stage, the variation of the PM velocities of OCs 
may be slight.
Based on that, the masses of progenitor clumps capable of producing OCs 
were obtained through a dynamical approach, whose statistics are 
concordant with the known results of Galactic clumps, such as the CMF.
In addition, as indicated by the masses of progenitor clumps,
the investigation confirms that many massive O-type stars exist
in present-day OCs, whose destructive stellar feedback can lead to a 
large number of embedded clusters being destroyed, even those
with sufficient densities can survive, and evolve to the phase of bound OCs.
These results could provide helpful indications of the OC formation and 
are expected to blaze a new trail for studying star formation in our Galaxy.
%

\normalem
\begin{acknowledgements}
We appreciate the anonymous referee for the comments which help us 
to improve the paper.
This work was funded by the NSFC grant No. 11933011 and by the Key Laboratory for Radio Astronomy. YJL thanks support from the Natural Science Foundation of Jiangsu Province (grant number BK20210999), the Entrepreneurship and Innovation Program of Jiangsu Province, and NSFC grant No. 12203104.
The authors acknowledge the open cluster catalogue compiled by \cite{cantat2020a}.
We used data from the European Space Agency mission \textit{Gaia} (\url{http://www.cosmos.esa.int/gaia}), processed by the \textit{Gaia} Data Processing and Analysis Consortium (DPAC; see \url{http://www.cosmos.esa.int/web/gaia/dpac/consortium}). Funding for DPAC has been provided by national institutions, in particular the institutions participating in the \textit{Gaia} Multilateral Agreement. 

\end{acknowledgements}

\bibliographystyle{raa}
\bibliography{reference}

\appendix 

\section{Additional Tables}
\label{app1}

Table~\ref{table:table_s1} presents the OCs harbouring O-type stars. 

\renewcommand\arraystretch{1.1}
\setlength{\tabcolsep}{0.6mm}
\begin{table}[!hbt]
\centering
\caption{The present-day OCs harbouring O-type stars.}
\label{table:table_s1}
\begin{tabular}{lll|lll} 
\hline\noalign{\smallskip} \hline 
OC name (${\rm log}_{\rm 10}({\rm Age})$) & $N$  & O name (Spectral type)  &
OC name (${\rm log}_{\rm 10}({\rm Age})$) & $N$  & O name (Spectral type)   
\\ \hline 
Alessi 43 (7.06) &  1 & HD 75759 (O9V+B0V) &
Berkeley 87 (6.92) &  1 & NGC 6913+37 513 (O9) \\  \hline
Hogg 15 (6.34) &  1 & HD 311884 (O5?V?) &
LP\_2179 (5.80) &  1 & HD 194094 (O8.5III) \\  \hline
IC 1396 (7.08) &  1 & HD 239729 (O9V) &
NGC 457 (7.32) &  1 & Cl* NGC 457 Hoag 5 (O9.5IV) \\  \hline
Stock 8 (7.16) &  1 & LS V+34 21 (O8) &
NGC 6871 (6.74) &  1 & HD 190864 (O6.5III(f)) \\  \hline
BH 121 (6.42) &  1 & HD 101298 (O6.5IV((f)))   &
NGC 6193 (6.71) &  1 & HD 150135 (O6.5V((f))z) \\  \hline
UBC 545 (7.55) &  1 & LS 3656 (O9:) &
Trumpler 15 (6.95) &  1 & Cl Trumpler 15 20 (O:)  \\  \hline
UBC 609 (6.75) &  1 & Sh 2-208 1 (O9.5V) &
UPK 169 (7.14) &  1 & HD 207538 (O9.7IV)  \\  \hline
IRAS02232+6138 (6.85) &  1 & BD+61 411 (O6.5V((f))z)&
FSR 0647 (6.46) &  1 & LS I +57 138 (O8Vz)  \\  \hline
FSR 0696 (6.82) &  1 & Sh 2-217 2 (O9.5V) &
Wit 1 (6.35) &  1 & HD 52266 (O9.5IIIn)  \\  \hline
NGC 6847 (8.70) &  1 & Sh 2- 97 7 (O9/B0V) &
UBC 404 (6.74) &  1 & BD+60 2635 (O6V((f))) \\  \hline 
Collinder 316 (6.68) &  1 & HD 152590 (O7.5Vz) &
Teutsch 127 (6.74) &  1 & BD+55 2722 B (O9.5V) \\  \hline 
IC 1590 (6.79) &  1 & HD 5005 D (O9.2V) &
He2020\_60 (8.30) &  1 & TYC 0170-1152-1 (O7) \\  \hline
UBC 266 (7.11) &  1 & HD 96622 (O9.2IV) &
UBC 267 (7.11) &  1 & HD 97848 (O8V) \\  \hline
UBC 633 (7.13) &  1 & LS 489 (O9V) &
Collinder 223 (8.20) &  1 & LS 1614 (O9.5Ib) \\  \hline
Loden 153 (6.30) &  1 & HD 91824 (O7V((f))z) &
Feinstein 1 (7.20) &  1 & HD 96670 (O8.5f(n)p) \\  \hline
Hogg 22 (6.80) &  1 & CPD-46 8221 (O9.7II-III) &
Havlen-Moffat 1 (6.95) &  1 & Cl HM 1 8 (O5III(f)) \\  \hline
IRAS20286+4105 (7.95) &  1 & V1827 Cyg (O6Iaf+O9:Ia:) &
Collinder 240 (6.95) &  1 & [J80] 1-123 (O8) \\  \hline
SAI 113 (6.95) &  1 & 2MASS J10224096-5930305 &
Majaess 133 (6.48) &  1 & 2MASS J10583238-6110565 \\ 
  &   &  (O7V((f))n) &
  &   &  (O5V((f))+O7V((f))) \\  \hline
SAI 24 (6.84)  &  1 & HD 18326  &
FSR 0236 (8.20) &  2 & [MT91] 227 (O9V) \\ 
  &   &  (O6.5V((f))z+O9/B0V:) &
  &   &  Cyg OB2 6 (O8.5V(n)) \\  \hline
Muzzio 1 (6.89) &  2 & CPD-47 2962 (O7V((f)))  &
NGC 1893(6.64) &  2 & LS V+33 15 (O7V(n)z) \\ 
  &   &  MO 2-56 (O9.5V) &
  &   &  BD+33 1025 A (O7.5V(n)z) \\  \hline 
Collinder 232 (6.30) &  2 & HD 303311 (O7V)  &
Pismis 20 (7.50) &  2 & [OM80] 40 (O9.5Ib) \\ 
  &   &  HD 93250 AB (O4IV(fc)) &
  &   &  Cl* Pismis 20 MV 2 (O8.5I) \\  \hline 
UBC 344 (6.54) &  3 & BD-12 4964 (O7V:)  &
NGC 2244 (7.10) &  3 & HD 46056 A (O8Vn) \\ 
  &   &  LS 4880 (O6V((f))) &
  &   &  HD 46223 (O4V((f))) \\   
  &   &  LS IV -11 8 (O8V) &
  &   &  HD 46485 (O7V((f))n) \\  \hline 
 Trumpler 16 (7.13) &  3 & HD 93204 (O6/7V) &
 Berkeley 59 (6.10) &  4 & BD+66 1673 (O5.5((f))(n))  \\ 
  &   &  Cl Trumpler 16 3 &
  &   &  BD+66 1674 (O9.7IV:) \\   
  &   &  (O9.5/B0.5V)  &
  &   &  BD+66 1675 (O7.5Vz) \\  
  &   &  Cl Trumpler 16 112 (O7V)  &
  &   &  TYC 4026-0424-1 (O7V((f))z)  \\ \hline
  Trumpler 14 (7.80) &  5 & HD 303312 (O9.5/B0V)  &
  NGC 3603 (6.00) &  4 & Cl* NGC 3603 Sher 57 (O3III(f)) \\ 
  &   &  Cl Trumpler 14 20 (ON8V) &
  &   &  Cl* NGC 3603 MDS 48 (O3.5If*) \\   
  &   &  Cl Trumpler 14 5 (O9:V) &
  &   &  Cl* NGC 3603 MDS 24 (O4IV(f)) \\ 
  &   &  HD 93128 (O3.5V((f))) &
  &   &  Cl* NGC 3603 Sher 23 (OC9.7Ia) \\ \cline{4-6}
  &   &  HD 93161 B (O6.5IV((f))) &
  Trumpler 24 (6.92) & 1  & HD 152559 (O9.5V)   \\ \hline 
  Havlen Moffat 1 (6.60) &  5 & Cl HM 1 12 (O6If) &
  Westerlund 2 (--) &  4 & Westerlund 2 MSP 182 \\ 
  &   &  Cl HM 1 20 (O9.5V) &
  &   &  (O4III/V((f)))  \\   
  &   &  Cl HM 1 18 (O7V((f))) &
  &   &  WR 20a (O3If*/WN6ha)  \\ 
  &   &  Cl HM 1 19 (O9V) &
  &   &   Westerlund 2 MSP 188 (O4III/V) \\ 
  &   &  Cl HM 1 9 (O9.7V) &
  &   &   Westerlund 2 MSP 171 (O4/5V) \\  \hline 
IC 1805 (6.88) &  5 & BD+60 499 (O9.5V)  &
vdBergh-Hagen 121 &  5 & HD 308813 (O9.7IV(n)) \\ 
  &   &  BD+60 501 (O7V((f))(n)z) &
  &   &  HD 101190 (O6IV((f))) \\   
  &   &  HD 15570 (O4If) &
  &   &  HD 101191 (O8V)  \\  
  &   &  HD 15629 (O4.5V((fc))) &
  &   &  HD 101223 (O8V) \\  
  &   &  BD+60 513 (O7Vn) &
  &   &  HD 101298 (O6.5IV((f))) \\   \hline    
NGC 6357 (--) & 7 & [N78] 46 (O7.5Vz)  &
NGC 6611 (6.33) &  8 & Cl* NGC 6611 ESL 029 (O8.5V) \\
  &   &  Cl Pismis 24 10 (O9V) &
  &   &  Cl* NGC 6611 ESL 017 (O9V) \\ 
  &   &  Cl Pismis 24 3 (O7.5:V) &
  &   &  HD 168075 (O6.5V((f))) \\
  &   &  Cl Pismis 24 2 (O5V((f))) &
  &   &  NGC 6611 222 (O7V((f))z) \\  
  &   &  Cl Pismis 24 16 (O7.5V)  &
  &   &  BD-13 4928 (O9.5V) \\  
  &   &  Cl Pismis 24 17 (O3.5III(f*)) &
  &   &  BD-13 4930 (O8.5V)  \\   
  &   &  Cl Pismis 24 13 (O6V((f))z) &
  &   &  HD 168137 (O8Vz) \\    
  &   &   &
  &   &  HD 168183 (O9.5III + B3/5III/V) \\ \hline
NGC 6231 (7.14) & 8 & HD 326331 (O8IV((f))n)  &
NGC 6231 (7.14) &  8 & HD 152199 (O9.7V) \\
  &   &  HD 152147 (O9.7Ib) &
  &   &  HD 152247 (O9.2III) \\ 
  &   &  HD 326329 (O9.7V) &
  &   &  HD 152249 (OC9Iab)  \\
  &   &  HD 152314 (O9IV) &
  &   &  HD 152248 AaAb (O7Iabf+O7Ib(f)) \\  \hline
\end{tabular}
\note{$N$ is the number of O-type stars in OCs.}
\end{table}
\clearpage

\end{document}